\documentclass[prl,twocolumn]{revtex4}  
\usepackage{graphics}

\begin{document}

\title{A detector for continuous measurement of ultra-cold atoms in real time}
\author{C. Figl, L. Longchambon$^*$, M. Jeppesen, M. Kruger$^\dagger$, H. A. Bachor, N. P. Robins and J. D. Close}

\affiliation{Australian Centre for Quantum Atom Optics, The Australian National University, Canberra, 0200, Australia.\\
$^*$present address, Laboratoire de Physique des Lasers, Universite Paris 13, Villetaneuse\\
$^\dagger$present address, Department of Physics, University of Missouri}
\email{cristina.figl@anu.edu.au}

\begin{abstract}
We present the first detector capable of recording high-bandwidth real time atom number density measurements of a Bose Einstein condensate.
Based on a two-color Mach-Zehnder interferometer, our detector has a response time that is six orders of magnitude faster than current detectors based on CCD cameras while still operating at the shot-noise limit.
With this minimally destructive system it may be possible to implement feedback to stabilize a Bose-Einstein condensate or an atom laser.
\end{abstract}

\maketitle

\noindent 
Experiments in Bose-Einstein condensation (BEC) of dilute alkali gases have yielded fascinating and valuable insights into modern physics, with countless avenues of research yet to be explored.
In these experiments, the atom-density distribution is usually measured by the interaction of atoms with light, either through the 
absorption of a probe laser, or through the phase shift that the atoms imprint
onto the light~\cite{ketterlephasecontrast,lincoln}.
All experiments to date on ground state alkali BECs have collected data with relatively slow CCD (charge coupled device) cameras.
Future experiments, such as applying feedback to stabilize an atom laser outcoupled from a BEC, require real-time detection without significant heating of the BEC.
The detection bandwidth has to cover DC to MHz, the timescale on which BEC dynamics occur.
These are the demands that such a detector has to meet.

\begin{figure}[b]
\centerline{
\scalebox{0.32}
{\includegraphics{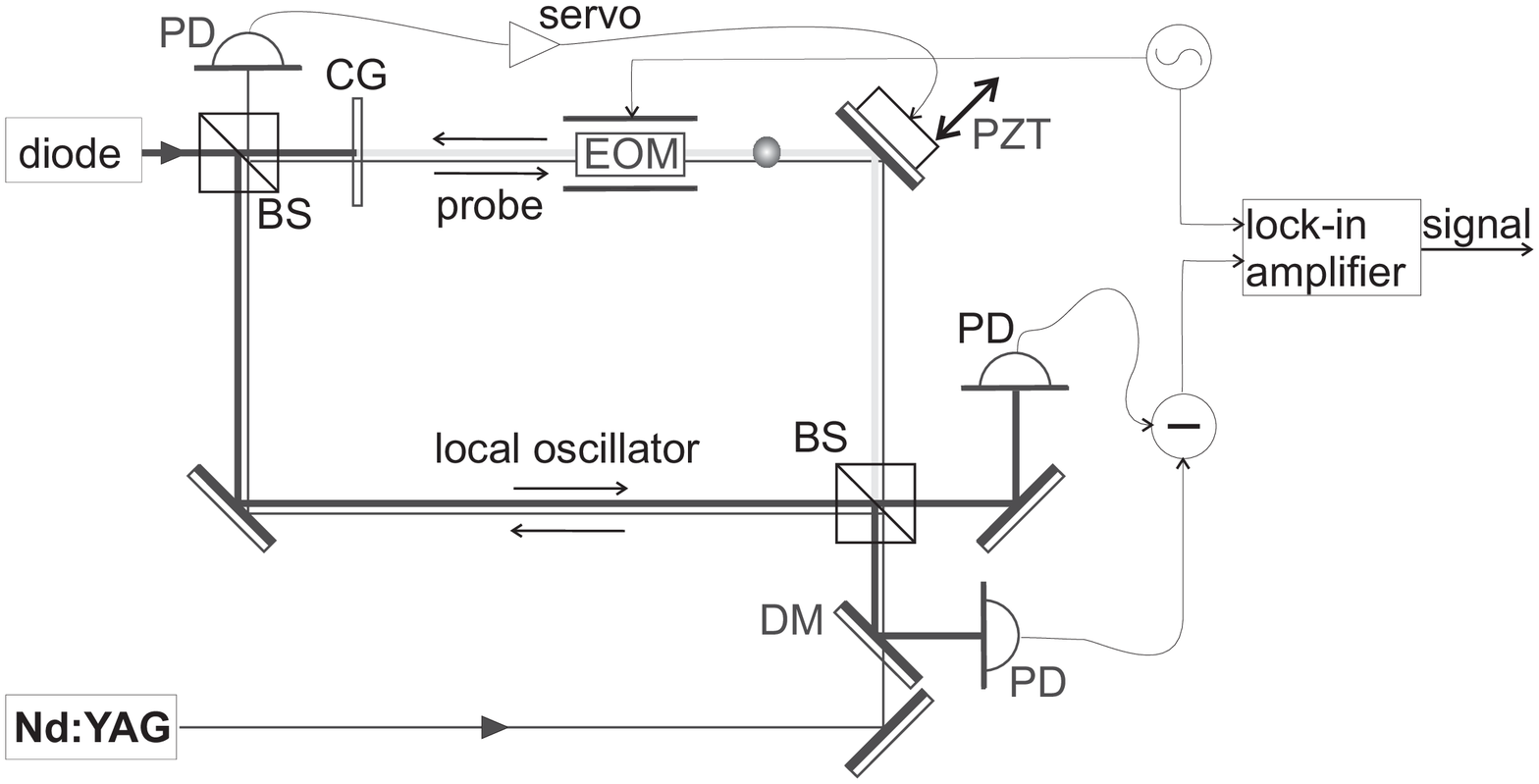}}}
\caption{Schematics of the setup. The beams of the two lasers are drawn separated for clarity. The circle indicates where atoms would be placed.
BS: 50-50 beamsplitters, PD: photodetector, CG: colour glass, DM: dichroic mirror
\label{fig.setup}}
\end{figure}

In this letter, we demonstrate and characterize such a minimally destructive, high-bandwidth, real time, and shot-noise limited detector based on an unbalanced two-color Mach-Zehnder interferometer (see Fig.~\ref{fig.setup}). 
It is designed to detect condensed $^{87}{\rm Rb}$ atoms but can be readily customized for other elements.
We expect this system to find applications in, for example, continuously measuring the formation of a BEC or collective oscillations of the condensate in feedback experiments.  
We envision that this technology is complementary to current imaging techniques as
our system does not supply spatial information, but rather temporal information about the peak density of a sample.  

We turn first to general design considerations.
The heating criterion imposes stringent limitations onto our system. With the kinetic temperature of a BEC of the order of 100\,nK, an atom scattering a single near-resonant photon is removed from the BEC due to the transfer of $\sim$500\,nK of recoil energy.
In the following, we consider a BEC of $10^5$ atoms confined to a volume of (10\,x\,10\,x\,50)\,$\mu{\rm m}^3$ with a lifetime of one second.
Thus, a scattering rate of $10^5$\,photons/second, or an absorbed light power $P_{\rm ab}$ of some ten femtowatts, is tolerable.

As discussed by Lye and coworkers~\cite{limits}, when detecting a fractional change in the column density $\delta \tilde n/\tilde{n}$, the best signal-to-noise ratio ($SNR_{\rm max}$) can be achieved in an interferometric measurement of the phase shift of the light $\phi = \tilde{n} \sigma_0 \Delta/(2(1+\Delta^2))$. $\tilde n$ is the column density, $\sigma_0 = 3 \lambda^2/(2 \pi)$ the resonant absorption cross section, and $\Delta$ the detuning of the light from the atomic resonance in multiples of half the atomic linewidth.
The atoms are placed in the probe arm of an interferometer with a local-oscillator arm around the atoms carrying much more light than the probe arm, and $\Delta$ should be large enough to reach the optically thin regime: 
$k = \tilde n \sigma_0/(1 + \Delta^2)\ll 1$.
Then,
\begin{equation}\label{eq.snr1}
SNR_{\rm max} = \frac{1}{2}\sqrt{\frac{\eta P_{\rm ab} \tilde n \sigma_0}{h\nu B}}\frac{\delta \tilde n}{\tilde n},
\end{equation}
where $B$ is the bandwidth, $h\nu$ the photon energy, and $\eta$ the quantum efficiency of the detector.
Setting $\eta = 1$, and $B = 1{\rm \,kHz}$, 
a full detection of the BEC, $\delta \tilde n / \tilde n = 1$, would be measured with a $SNR_{\rm max}$ of 85, hence, with $SNR = 1$, a minimum $\delta \tilde n/\tilde{n}$ of around 1\% could be detected. 

Notably, if we hold the photon absorption rate and background 
column density fixed, the $SNR$
depends only on the bandwidth of the detection system. 
No amount of detuning from atomic resonance, enhancement of refractive 
index via coherences in multi-level systems, or otherwise can improve on a 
simple measurement based on two-level atom physics~\cite{limits2}.
The only possible improvements would be a cavity around the atoms (enhancing the $SNR$ by the square root of the finesse) or the use of squeezed light (enhancing the $SNR$ by the squeezing factor). Both possibilities, however, constitute severe experimental difficulties and are not pursued here.

A Mach-Zehnder interferometer is susceptible to changes in its armlength.
Single-beam frequency modulation spectroscopy~\cite{alain} or a Sagnac interferometer are intrinsically immune to geometrical misalignment.
However, they suffer from the fact that the local-oscillator beam traces the same path as the probe beam and contributes to the destruction of the atomic cloud.
A demonstrated offset-Sagnac interferometer~\cite{jesoffset} seems to provide both separated beams and stability, but it was not operating at the shot-noise limit, and the maximum separation achievable (a few millimeter) is not enough to avoid scattering of light from the local oscillator into the BEC. 
In our Mach-Zehnder interferometer, the beam separation is large enough to pass the local-oscillator beam around the vacuum chamber containing the BEC.
However, vibrations due to laboratory  air conditioners, fans and seismic occurring below some kilohertz have to be suppressed.
We achieve this by locking the interferometer with a Nd:Yag laser.
Its wavelength of 1064\,nm is terahertz detuned from the megahertz wide atomic resonance of $^{87}{\rm Rb}$ at 780\,nm.
Being insensitive to the presence of the atoms, it is used to hold the interferometer at a fixed geometrical pathlength difference using a piezo translator supporting one mirror.
A probe laser with a frequency close to the atomic resonance records the atomic signal.
The probe and locking beam counterpropagate inside the interferometer and are overlapped carefully so that exactly the same optical path, including changes of the refractive index of the air, is probed. 
Massive optical mounts, and acoustic and thermal isolation provide a passive stabilization.
The remaining phase noise or locking-limited sensitivity of the locked interferometer
is $2\cdot 10^{-4}{\rm \,rad}/\sqrt{\rm Hz}$, measured, as all data in this paper,  over 1\,s with a bandwidth of 1\,kHz.

The requirement on the performance of the locking loop depends on the detuning of the probe laser.
The $SNR$ of Eq.~\ref{eq.snr1} is independent of detuning, assuming only that it is large enough so that the BEC is optically thin ($k \ll 1$).
We can operate our probe laser either near detuned at low power with a poor phase sensitivity (and a large atomic phase shift) or far detuned at high power with a good phase sensitivity.
Both schemes, in theory, would provide the same $SNR$ at the same level of heating.
In practice, we want to operate where the phase shift due to the atoms is as large as possible to relax the locking requirement, i.~e. as close to resonance as possible, while still operating in the optically thin regime. 
In our example, $k = 0.01$ results in a detuning of $\Delta = 170$, thus $P_{\rm Pt} = P_{\rm ab}/(1 - e^{-k}) =$ 2.6\,pW of light can be sent through the atoms.
The full BEC would be detected with a phase shift of around one radian, corresponding to a sensitivity of $3.2\cdot 10^{-4} {\rm rad}/\sqrt{\rm Hz}$. Thus, the sensitivity is not limited by the locking loop. 

We turn now to a more detailed description of our setup. The probe beam at 780\,nm is provided by a commercial external cavity diode laser locked 
to the $^{87}{\rm Rb}$\,$S_{1/2}(F=2)\rightarrow $\,$ P_{3/2}(F=2,3)$ crossover using saturated absorption spectroscopy.
The signal is detected at the interferometer outputs by photodiodes. 
The use of photodiodes instead of CCD cameras allows for the high bandwidth.
The Si-PIN photodiodes (BPX 65) are placed in custom low-noise DC coupled transimpedance detectors~\cite{pd}.
At 780\,nm, the photodetectors have a quantum efficiency of 0.72 with a noise equivalent power of $7$\,pW/$\sqrt{\rm Hz}$.
This allows for shot-noise sensitivity in the frequency range from 0.5 to 10\,MHz for optical powers from $70$\,$\mu$W to $3$\,mW. 
We operate our photodiodes at the shot-noise limit with the $\sim$\,mW local oscillator.
The laser intensity in the probe arm (where the BEC is to be placed) is attenuated with Schott color glass filters (RG 830) to $\sim$\,pW leaving the 1064\,nm locking laser intensity nearly unaffected.
We detect both output ports of the interferometer and subtract the corresponding signals (balanced homodyne detection).
The (anticorrelated) signal is thus enhanced by a factor of 2 while the noise enhancement is only $\sqrt{2}$ because the (uncorrelated) noise terms add in quadrature.
Apart from the enhancement of the $SNR$, homodyne detection suppresses (in our case by 50\,dB) correlated noise such as laser intensity noise.

Photodetectors and other electronic components commonly show a $1/f$ noise making direct shot-noise limited measurements in the acoustic band difficult.
We phase-modulate one beam with an electro-optic modulator (EOM) at $\omega_{\rm m}=2.5{\rm \,MHz}$ to transfer the low-frequency signal to a frequency range in which the electronic noise is low and then demodulate the detected signal. 
Thus, two sidebands at $\pm \, \omega_{\rm m}$ relative to the carrier frequency are created, each of power $J_1(m)^2$ where $J_1$ is the first order Bessel function of first kind and $m$ the modulation depth.
Placing the EOM in the highly attenuated probe beam reduces the background from residual amplitude modulation.
The signal is demodulated using a lock-in amplifier (SRS SR844)
and recorded on an oscilloscope.
The detection bandwidth is set by the choice of the low-pass filter in the lock-in amplifier.

The demodulated rms photocurrent $i_{\rm S}$ is 
\begin{equation}\label{eq.signal}
i_{\rm S} =\pm \frac{\alpha \eta e}{\sqrt{2} h\nu}  \sqrt{P_{\rm LO} P_{\rm Pt}}J_1(m) \sin{(\phi - \gamma)}, 
\end{equation}
where the sign depends on the output port.
$P_{\rm Pt}$ ($P_{\rm LO}$) is the transmitted light power of the probe (local oscillator) beam. 
$\gamma$ comprises phase shifts from all sources except for the atomic sample and defines the operating point of the interferometer.
$e$ is the electron charge, and $\alpha$ includes all electronic amplification and losses.
We assumed $\Delta \gg \omega_{\rm m}$ so that the absorption and phase shift of the probe-beam carrier and its sidebands can be approximated to be equal.
Eq.~\ref{eq.signal} describes the probe-beam sidebands beating with the strong local oscillator (terms of the order of the weak probe-beam power $P_{\rm Pt}$ were neglected).
The signal is detected on a shot-noise background of 
\begin{equation}\label{eq.shot}
i_{\rm shot} = \frac{\alpha e }{\sqrt{2}} \sqrt{\frac{ \eta  P_{\rm LO} B}{2 h \nu}}. 
\end{equation}
The $SNR$ of our detection technique is thus 
\begin{equation}\label{eq.snr}
SNR = 2 J_1(m) \sqrt{\frac{ \eta P_{\rm Pt} }{h\nu B}} \delta \phi 
    =  J_1(m) \sqrt{\frac{ \eta  P_{\rm ab} \tilde n \sigma_0 }{h\nu B}}\frac{\delta \tilde n}{n}
\end{equation}
assuming $\Delta \gg 1$, $\delta \phi \ll 1 $ with $(\phi - \gamma)$ close to zero, and $k \ll 1$.
A comparison to Eq.~\ref{eq.snr1} shows that the mod\-u\-la\-tion, while making the measurement shot-noise limited, comes at the price of decreasing the $SNR$ by $2 J_1(m)$, 0.51 in our case. 
A heterodyne technique by using acousto optical modulators in the interferometer to shift one beam in frequency with respect to the other one would allow to reach $SNR_{\rm max}$. 
Choosing $\omega_{\rm m} \gg \Delta$ would serve the same purpose and would allow to make the interferometer less susceptible to vibrations~\cite{jessthesis}. 
However, both possibilities come with large technical noise making operation at the shot-noise limit very unlikely.

We determined the noise of the local-oscillator beam at 2.5\,MHz as a function of $P_{\rm LO}$ by calculating the standard deviation of a measured trace.
The results agree within 16\% with calculations from Eq.~\ref{eq.shot}, and 
a log-log fit to the data results in a slope of $0.51\,\pm\, 0.02$. 
Shot-noise is the only noise source that scales with the square root of the light power. 
A measurement with the interferometer locked near mid fringe
proves that there is no detectable contribution of other noise sources:
the fits to both sets of data lie within each other's prediction bounds. 
This shows that our system is indeed shot-noise limited.

\begin{figure}[b]
\centerline{
\scalebox{0.32}{
\includegraphics{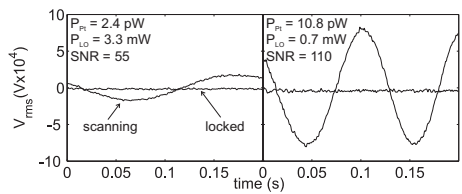}}}
\caption{Signal and noise at 2.5\,MHz. Light powers and $SNR$ as given in the graphs.
\label{fig.trace}
}
\end{figure}
\begin{figure}[tb]
\centerline{
\scalebox{0.3}{
\includegraphics{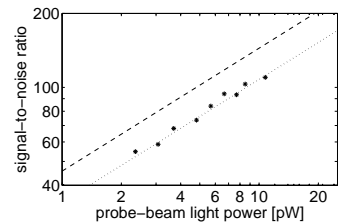}}}
\caption{$SNR$ as a function of the probe-beam light power: measurement ($*$), calculation including ($\cdot\cdot$)/without (- -) losses and inaccuracies of calibration as given in the text. 
\label{fig.snr}
}
\end{figure}

In order to determine the $SNR$, we mimic an atomic phase shift by scanning the mirror with the piezo (see Fig.~\ref{fig.trace}). Fitting a sine curve to the data, we calibrate the voltage dependence on the phase shift for a range of input light powers.
With the measured noise level we determine the $SNR$ for a $\pi$ phase shift (Fig.~\ref{fig.snr}) and compare it to calculations from Eq.~\ref{eq.snr} (dashed line in Fig.~\ref{fig.snr}).
The scaling of the data (a fit has a slope of $0.49 \pm 0.06$) is in excellent agreement with Eq.~\ref{eq.snr}. 
The theoretical $SNR$ is, however, about 30\% larger than measured.
In Eq.~\ref{eq.snr}, losses of the probe beam from mirrors and beamsplitters as well as imperfect mode-matching of the two interfering beams were not taken into account. Together with inaccuracies in the determination of amplification and losses in the detection chain this accounts for the deviation.
The dotted line in Fig.~\ref{fig.snr} are calculations from Eq.~\ref{eq.snr} with these deviations taken into account as a scaling factor.
The current setup is limited by the interferometer locking-loop to a $SNR$ of around 300.

With the presented data, we can characterize the performance of the system:
In the BEC we considered, with a bandwidth of 1\,kHz and a heating rate of one photon/atom/second, we would be able to track the fractional change in column density with a sensitivity of $0.09\%/\sqrt{\rm Hz}$.
Thus, we have succeeded in building a shot-noise limited minimally destructive device capable of tracking the dynamics of a BEC with a high bandwidth in real time.

We acknowledge important discussions with the gravity wave and the quantum optics group at the Australian National University.
CF aknowledges funding by the Alexander von Humboldt foundation.
This work was finacially supported by the Australian Research Council.

\end{document}